\documentclass[conference]{IEEEtran}
\usepackage{booktabs} 

\usepackage{subcaption}

\usepackage{graphicx}

\usepackage{adjustbox}

\usepackage{hyperref}
\hypersetup{
    colorlinks=true,
    linkcolor=blue,
    filecolor=magenta,      
    urlcolor=cyan,
}
\usepackage{listings}
\usepackage{color}
\usepackage{xcolor}
 
\definecolor{codegreen}{rgb}{0,0.6,0}
\definecolor{codegray}{rgb}{0.5,0.5,0.5}
\definecolor{codepurple}{rgb}{0.58,0,0.82}
\definecolor{backcolour}{rgb}{0.95,0.95,0.92}
 
\lstdefinestyle{mystyle}{
    backgroundcolor=\color{backcolour},   
    commentstyle=\color{codegreen},
    keywordstyle=\color{magenta},
    numberstyle=\tiny\color{codegray},
    stringstyle=\color{codepurple},
    basicstyle=\footnotesize,
    breakatwhitespace=false,         
    breaklines=true,                 
    captionpos=b,                    
    keepspaces=true,                 
    numbers=left,                    
    numbersep=5pt,                  
    showspaces=false,                
    showstringspaces=false,
    showtabs=false,                  
    tabsize=2
}
 
\lstdefinelanguage{swift}
{
  morekeywords={
    func,if,then,else,for,in,while,do,switch,case,default,where,break,continue,fallthrough,return,
    typealias,struct,class,enum,protocol,var,func,let,get,set,willSet,didSet,inout,init,deinit,extension,
    subscript,prefix,operator,infix,postfix,precedence,associativity,left,right,none,convenience,dynamic,
    final,lazy,mutating,nonmutating,optional,override,required,static,unowned,safe,weak,internal,
    private,public,is,as,self,unsafe,dynamicType,true,false,nil,Type,Protocol,
  },
  morecomment=[l]{//}, 
  morecomment=[s]{/*}{*/}, 
  morestring=[b]" 
}

\definecolor{keyword}{HTML}{BA2CA3}
\definecolor{string}{HTML}{D12F1B}
\definecolor{comment}{HTML}{008400}

\begin{document}
%
\title{On Evaluating the Effectiveness of the HoneyBot: A Case Study}

\author{\IEEEauthorblockN{Celine Irvene}
\IEEEauthorblockA{Georgia Institute of Technology\\
cirvene3@gatech.edu}
\and
\IEEEauthorblockN{David Formby}
\IEEEauthorblockA{Georgia Institute of Technology\\
djformby@gatech.edu}
\and
\IEEEauthorblockN{Raheem Beyah}
\IEEEauthorblockA{Georgia Institute of Technology\\
rbeyah@gatech.edu}}


%


\IEEEoverridecommandlockouts
\makeatletter\def\@IEEEpubidpullup{6.5\baselineskip}\makeatother
\IEEEpubid{\parbox{\columnwidth}{
    Workshop on Usable Security (USEC) 2019 \\
    24 February 2019, San Diego, CA, USA \\
    ISBN 1-891562-57-6 \\
    https://dx.doi.org/10.14722/usec.2019.23xxx \\
    www.ndss-symposium.org
}
\hspace{\columnsep}\makebox[\columnwidth]{}}

\maketitle

\begin{abstract}
In recent years, cyber-physical system (CPS) security as applied to robotic systems has become a popular research area. Mainly because robotics systems have traditionally emphasized the completion of a specific objective and lack security oriented design. Our previous work, HoneyBot \cite{celine}, presented the concept and prototype of the first software hybrid interaction honeypot specifically designed for networked robotic systems. The intuition behind HoneyBot was that it would be a remotely accessible robotic system that could simulate unsafe actions and physically perform safe actions to fool attackers. Unassuming attackers would think they were connected to an ordinary robotic system, believing their exploits were being successfully executed. All the while, the HoneyBot is logging all communications and exploits sent to be used for attacker attribution and threat model creation. In this paper, we present findings from the result of a user study performed to evaluate the effectiveness of the HoneyBot framework and architecture as it applies to real robotic systems. The user study consisted of 40 participants, was conducted over the course of several weeks, and drew from a wide range of participants aged between 18-60 with varying level of technical expertise. From the study we found that research subjects could not tell the difference between the simulated sensor values and the real sensor values coming from the HoneyBot, meaning the HoneyBot convincingly spoofed communications. 
\end{abstract}



%

\section{Introduction}\label{sec:introduction}
Often robotic systems come in different shapes, sizes, and colors. At the end of the day the thing they all have in common is that they are built for a purpose. Whether it be for performing telesurgery \cite{tozal}, automating smart factories \cite{saribatur}, or assisting infantry soldiers \cite{srin}, robots are built for accomplishing specific objectives. In the past and even at the present, when designing robots to achieve these objectives, roboticists often exclude security principles and techniques. These exclusions are usually the manifestation of having never been formally trained on secure practices or the result of some implicit hardware/software constraints of their system. These security shortcomings in robotic system have led to our two main research questions: when put in a situation to take advantage of a robotic system, will users do so and of those users who will, can they detect the difference between manipulating the real system and a simulated one?

\subsection{HoneyBot for Robotic Systems}
With robotic systems becoming more and more integrated into the fabric of everyday life, it is paramount to secure them before they become safety hazards to society. In our previous work \cite{celine}, we proposed HoneyBot, the first software hybrid interaction honeypot specifically designed for networked robotic systems. 

\begin{figure}[h]
\centering
\includegraphics[scale=.27]{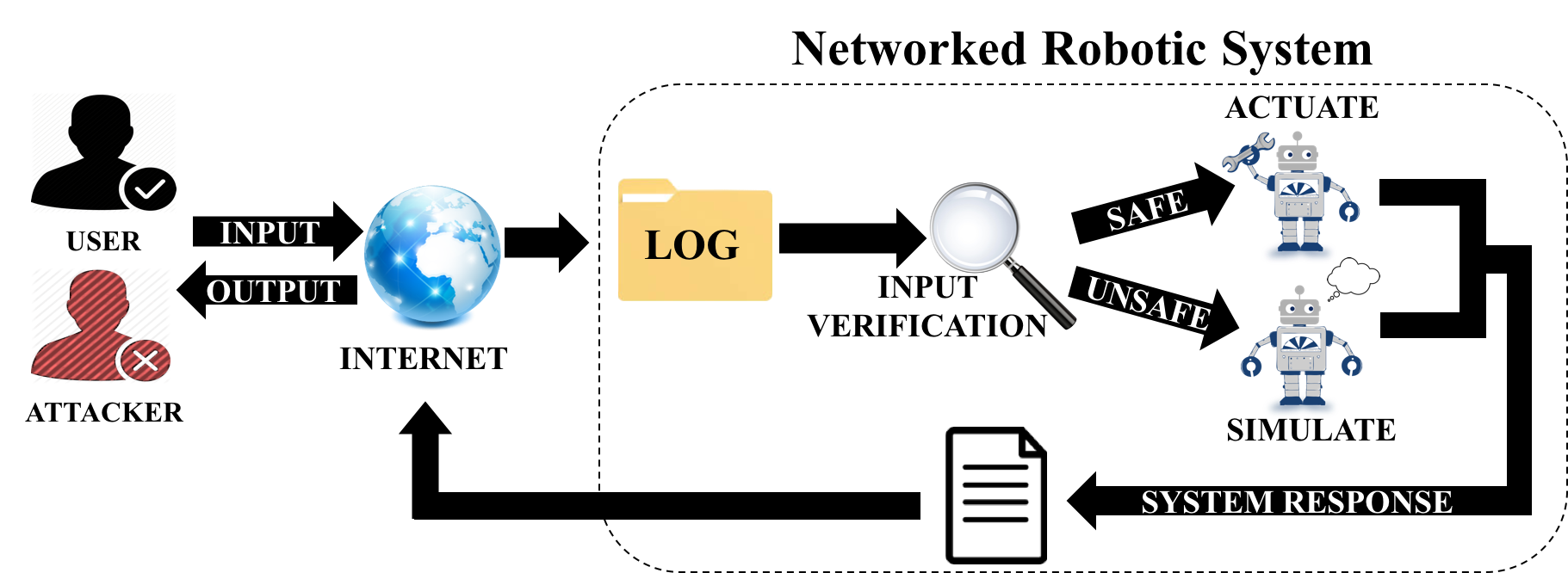}
\caption{HoneyBot system architecture.}
\label{sysarch}
\end{figure}

HoneyBot is a hybrid interaction honeypot that alternates between simulation and physical actuation. It takes into account device physics and uses device modeling to provide realistic simulations for requested commands when they are deemed too hazardous, for either the robot or the environment, to be performed. If the requested command is deemed benign or otherwise safe, it is physically performed by the robotic system. In both cases, whether a command is deemed safe or unsafe the system response is sent back to the attacker as depicted in Figure \ref{sysarch}. 

\subsection{HoneyBot User Evaluation}
\begin{figure}[h]
\centering
\includegraphics[scale=.3]{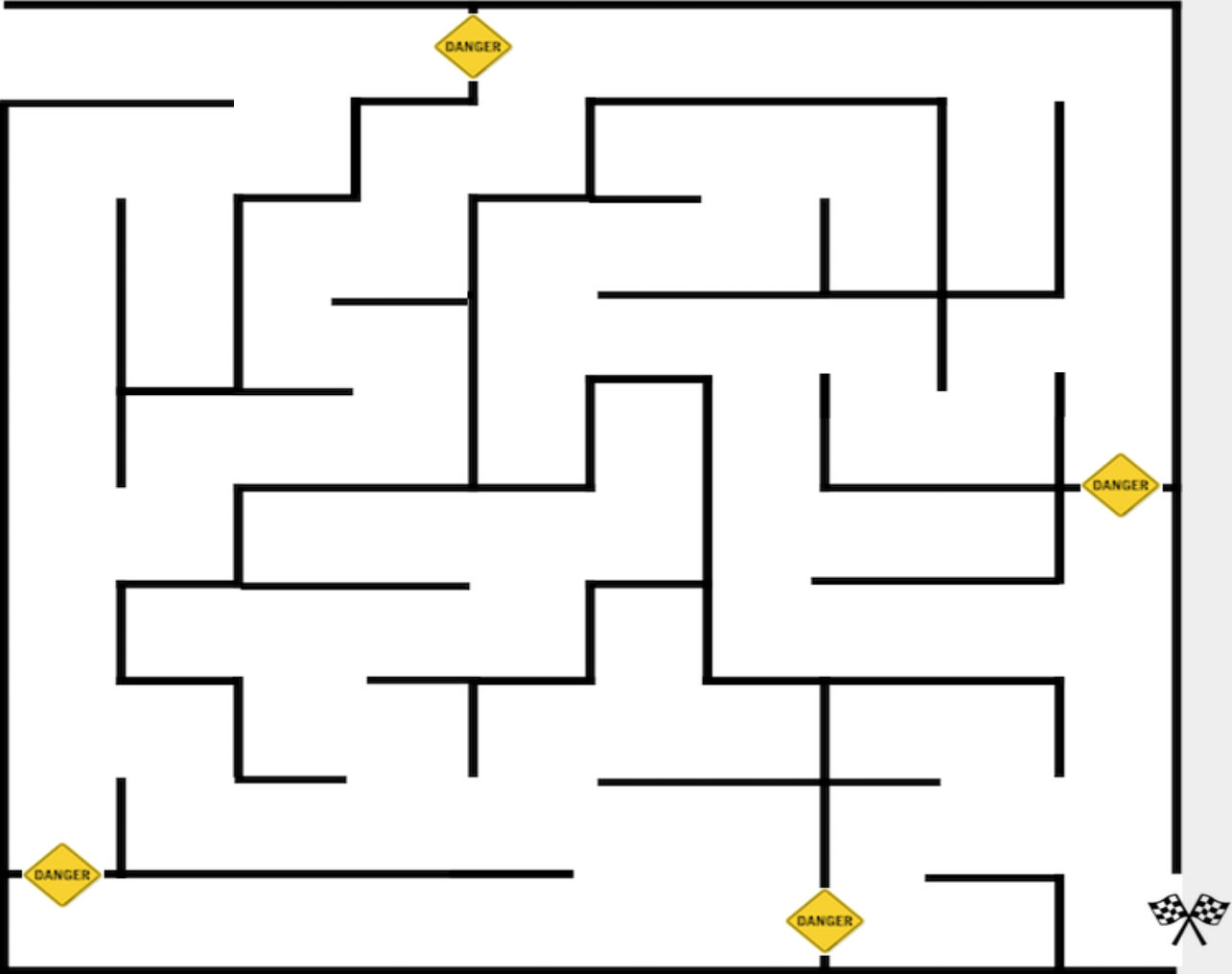}
\caption{Online HoneyMaze with danger signs throughout.}
\label{onlinemaze}
\end{figure}

To evaluate the HoneyBot, we obtained IRB approval from the Georgia Institute of Technology and employed a longitudinal study over the course of several weeks with 40 recruited participants. The users connected to the HoneyBot via a web GUI and were given access to remotely navigate the robot through a maze, shown in Figure \ref{onlinemaze}. Participants were told that their goal was to control a robot through an assessment course for testing the navigational capabilities of the remotely accessible robot under various constraints for determining the optimal constraint profile for the performance and efficiency of the robot. They were informed that their actions on the web GUI would cause a robot to physically move through a real maze and instructed to use the arrow keys on their keyboard in the online virtual interface to navigate the GUI robot as quickly as possible through the online maze to the finish flags. Participants were also instructed to use the sensor values on the online control panel (located to the right of the online virtual maze) to maintain situational awareness of the robot. In order to add a sense of urgency to the users there was a 60 second time constraint placed on the navigation task, but participants were allowed a 75 second preview of the maze to plan their route. This time limit was enforced to make the situation more akin to how a real attacker would act after gaining access to a computer system. Usually, they aim to perform their malicious payload as quickly as possible then exit the system. The 75 second preview can be thought of as an abbreviated reconnaissance phase before the actual cyber attack. This is the time when an attacker would analyze the network or system to determine its weakest points and/or make a plan for carrying out the payload. This preview/reconnaissance phase is what allows the actual task completion to be so short, once the plan is made carrying it out is simple. After completing the experiment, participants completed a short survey about their experiences. The survey was crafted in such a way to determine what navigational routes participants took through the maze, why they chose the routes they did, whether or not they completed the maze in the given time, and what, if anything, did they notice about the robot's sensor control panel.

What the participants of the research study didn't know was that in reality they were only controlling the robot through the real maze part of the time. On the online maze there were four paths marked with danger signs, and all users were given the same instructions regarding the viable paths to take. They were simply told, "consider all possible routes". These danger signs indicated no real danger to the robot, but were instead symbolic of a "restricted zone" on a real computer system. Given that the HoneyBot is a honeypot for robotic systems, the danger signs served as this honey, or temptation to go outside of the "safe zone". The danger signs marked shortcuts through the maze, and if the participants attempted to go near them the robot would bust through the sign emerging on the other side. The online maze was setup in such a way that the only way to complete the maze in the time given was to "take the honey" and cut through at least one danger sign. 

We have implemented our proof of concept HoneyBot in a hardware prototype programmable ground robot. Our user study shows that the majority of users cannot detect the difference between actually controlling the HoneyBot and the HoneyBot simulating control of the system sending back "spoofed" system responses. We found that on average 35\% of users will deviate or make riskier choices in the presence of pressure to experience higher reward. 
In summary, the main contribution of this work is the evaluation of the effectiveness of a HoneyBot prototype via a user study. Our results show that users are unable to determine the difference between physically controlling the HoneyBot versus the HoneyBot simulating control of the system.

The rest of this paper is organized as follows. In Section 2 we present related work in the area of honeypots, Section 3 describes the proof of concept HoneyBot design and implementation in detail, Section 4 details the HoneyBot experimental design, Section 5 discusses the experimental results of the user study. Section 6 and 7 discuss our conclusions and future work.

\section{Related Work}
Until recently, honeypots have generally been tools used only in IT networks to both detect attackers infiltrating the network, and to monitor their behavior and learn their attack strategies. The fidelity of these honeypots has ranged from low interaction to high interaction, and their effectiveness has been evaluated on the basis of how easily and accurately attackers can detect that they are in a honeypot using automated techniques. Low interaction honeypots are easily detected while high interaction honeypots consisting essentially of real systems are significantly harder to detect. However, these techniques that are used to evaluate traditional honeypots fail to measure the effectiveness of high-fidelity cyber physical system (CPS) honeypots, where not only must the software behave like a real system, but the reported physics of the system must also. Automatically determining the difference between a high-fidelity physical simulation and the real physical process is very difficult, so attackers must be able to subjectively make the decision using their own human intuition. Therefore, novel methods for evaluating the effectiveness of CPS honeypots that take this factor into account are necessary.

The first CPS honeynet addressed supervisory control and data acquisition (SCADA) networks and was created by Pothamsetty and Franz of the Cisco Infrastructure Assurance Group (CIAG) in 2004 \cite{pothamsetty}. The researchers were able to simulate popular PLC services with the goal to better understand the risks of exposed control system devices. This work laid the foundation for many other CPS honeypots \cite{rist, wilhoit}, including our own previous work creating a framework for hybrid interaction CPS honeypots \cite{honeyphy} and honeypots for robotic systems \cite{celine}. However, the fidelity of these hybrid interaction CPS honeypots were only evaluated by visually comparing the simulations to real values, and not testing whether a true adversary could tell the difference.

Security has always been an arms race between attackers and defenders, and honeypot detection is no exception. Attackers are constantly discovering new combinations of evidence that fingerprint a honeypot's identity and researchers are endlessly trying to modify them to blend in. For example, an early high-interaction honeynet, Sebek \cite{sebek}, which was exposed the very next year as a honeynet by techniques described in \cite{nosebreak}. The next evolution of Sebek, called Qebek, attempted to hide more effectively by using virtualized high-interaction systems \cite{qebek}. For many non-high-interaction honeypots, evasion boils down to finding the edges of emulation for the presented services, but timing approaches have also been used \cite{defibaugh}. For example, Kippo is a popular medium-interaction honeypot for the SSH (Secure Shell) service \cite{kippo}. Kippo is easily detected by sending a number of carriage returns, and noting the output difference from production SSH servers \cite{kippo-detect}. While high interaction honeypots are harder to detect, many rely on virtualization. Virtualization technologies usually leave their own fingerprints, such as device names, device driver names, file-system hallmarks, and loaded kernel modules \cite{holz}. Even though these fingerprints can be altered, there exist a rich set of techniques for detecting virtualization (and defeating detection attempts) from the malware-analysis field \cite{chen}. 

Other, more honeypot-technology agnostic, detection techniques have been proposed. Some of these techniques rely on the liability issues inherent in hosting deliberately compromised machines. A botnet architecture proposed in \cite{zou} leverages the honeypot owner's desire to restrict outgoing malicious traffic to authenticate new hosts before integrating them into the botnet. Specifically, the new host is directed to send apparently malicious traffic to an already compromised "sensor." Most honeypot systems will attempt to identify and block or modify this malicious traffic, so whether the sensor receives the traffic unaltered can be used to determine if the new host is genuine. This work was built upon in \cite{hayatle}, where multiple pieces of evidence can be formally combined to derive a metric of likelihood that a host is a honeypot. This evidence could be the virtualization status of the host, the diversity of software on the host, the level of activity of the host, or the difficulty in compromising the host. This newer technique is presented in the context of a botnet, but the generalized belief metric is equally applicable to any honeypot technology, depending on the evidence used. There are also techniques, described in two more recent surveys \cite{bringer} \cite{nawrocki}, which elaborate on the ideas above, including finding edges of emulation, finding subtle discrepancies that indicate virtualization, or analyzing the results of communication with an already compromised sensor. 

Since the physics of every CPS system is unique, it is much more difficult for attackers to create automated tools for detecting physical simulations. Therefore, the goal of a CPS honeypot is to be realistic enough to fool a human attacker's intuition of the physics of the process, which most closely resembles social engineering and phishing attacks. The effectiveness of these kinds of attacks have been extensively studied \cite{why_phishing, phil, phinding_phish}, but there are significant differences between fooling a civilian with a phishing email and fooling an attacker with a physical simulation. This work employs real human subjects to test the fidelity of the CPS honeypot physical simulation, which was not done in our previous work \cite{celine}. The high-fidelity, hybrid interaction HoneyBot system was deployed in a maze environment and human subjects were asked to remotely navigate the robot through the maze and later questioned on how real the challenge seemed.

\section{Proof of Concept HoneyBot}
In order to facilitate the most representative user evaluation of the HoneyBot as possible, we constructed the HoneyBot framework and architecture on a real robotic system. The design details of this proof of concept HoneyBot and the subsequent user experimentation are described in this section.

\begin{figure}[ht!]
    \centering
    \begin{subfigure}[ht!]{.12\textwidth}
        \centering
        \includegraphics[scale=.02]{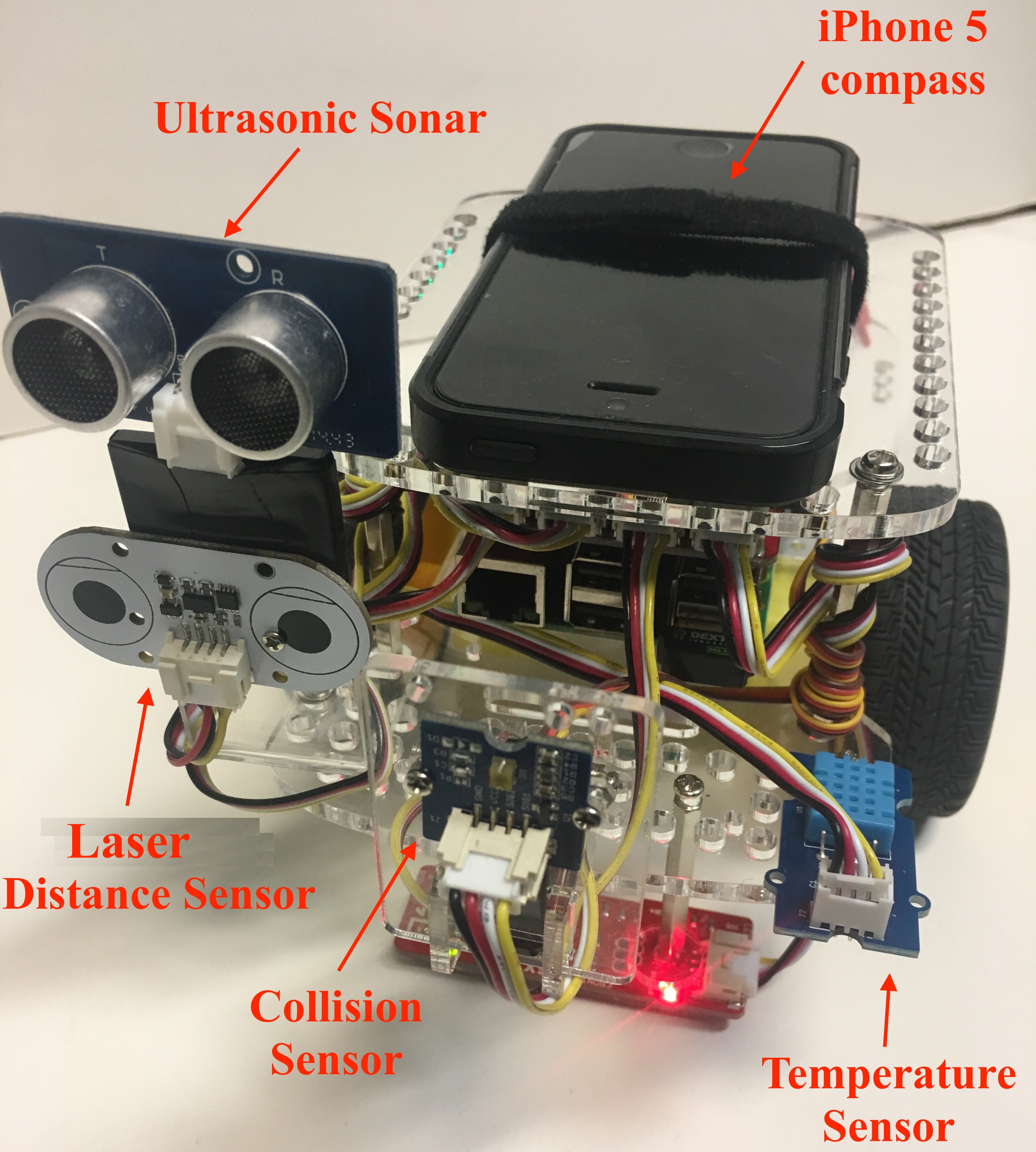}
        \caption{}
        \label{gpg3}
    \end{subfigure}
    ~ 
    \begin{subfigure}[ht!]{.12\textwidth}
        \centering
        \includegraphics[scale=.2]{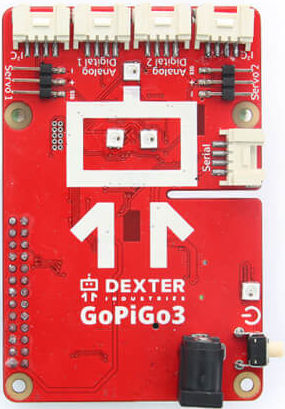}
        \caption{}
        \label{gpg3 board}
    \end{subfigure}
        ~ 
    \begin{subfigure}[ht!]{.12\textwidth}
        \centering
        \includegraphics[scale=.25]{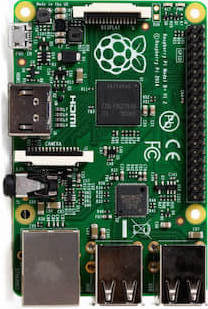}
        \caption{}
        \label{rpi}
    \end{subfigure}
    \caption{Images of the (a) fully outfitted GoPiGo3 Robot Car (b) GoPiGo3 Circuit Board (c) and Raspberry Pi 3.}
    \label{robot components}
\end{figure}

The GoPiGo 3 \cite{gopigo}, shown in Figure \ref{gpg3}, was the chosen robotic system for the proof of concept HoneyBot. This platform was selected because of the ease of programming, through its support of the Python programming language, and the many I/O interfaces for attaching various robotic sensors. In addition to this the GoPiGo 3 was selected over the GoPiGo 2, used for initial model development, because of its magnetic encoders which ensure accurate robot control and its redesigned power management system which gives it longer battery life. These upgrades were crucial to performing the evaluation described in Section \ref{experiment design}. The GoPiGo 3 Robot Car is a ground robot that consists of six major components: a GoPiGo 3 circuit board, a Raspberry Pi 3 \cite{rpi}, two motors, two wheels, various sensors, and a battery pack. The GoPiGo 3 circuit board, shown in Figure \ref{gpg3 board}, can be considered the secondary controller of the GoPiGo 3 Robot Car. It connects to the header pins of the Raspberry Pi 3, shown in Figure \ref{rpi}, and receives motor control commands as well as provides status updates about the various connected sensors. The Raspberry Pi 3 is the main controller of the robot and can be accessed via direct connection (through its HDMI port), SSH, or VNC. The Raspberry Pi 3 runs the Raspbian OS, a version of Linux created especially for single board computers.  

\subsection{HoneyBot Software}
The HoneyBot software was written in Python 2.7 and is made up of three main modules: the Robot Web Server, The Robot Controller, and the HoneyBot Module.

\subsubsection{Robot Web Server}
The robot web server is essentially the \textit{Internet Interface Module} from the HoneyPhy framework and serves to communicate and transport commands from the front end (web page) to the robot's actual hardware. The server was written using the Tornado web framework \cite{tornado}. The web server is the process that is called to spin up every other module. When executed the web server instantiates a robot object (the Robot Controller), a HoneyBot object (the HoneyBot Module), serves up the HoneyBot login page, and facilitates all web requests from clients through web sockets. The HoneyBot login page was used to safeguard the robot experimentation and evaluation process by defending access to the robots hardware with a rotating pairs of usernames and passwords. Before anyone could access the robot experiment website they had to enter a correct username/password pair and each set of credentials could only be used once before being invalidated, like a nonce.

\subsubsection{Robot Controller}
The robot controller can be considered the \textit{Process Model} from the HoneyPhy framework \cite{honeyphy} as it receives commands from the robot web server and translates them to navigational commands for the robot to perform or simulate. For instance, if the user clicks the right arrow key this is transported over a web socket from the client web page to the Tornado web server backend. The backend makes a call to the robot controller object which converts it to a navigation command and passes that to the Input Verification Module along with the robots' current status. The Input Verification Module then determines whether or not the command is safe to perform and if it is it gets sent to the robot's motors. If unsafe the HoneyBot Module queries the sensor \textit{Device Models} and spoofed data is returned. 





            

\subsubsection{HoneyBot Module}
The honeybot module is responsible for running a background process that constantly queries the robot for true sensor data. If the Input Verification Module detects an unsafe command the robot controller will call the honeybot modules' simulateStatusUpdate method and each of the robot's sensor \textit{Device Models} will be queried for simulated data. The simulated data was collected through empirical observation and is described in detail in Section \ref{sensormodels}.

\subsection{HoneyBot Sensors}
\begin{figure*}[t!]
    \centering
    \begin{subfigure}[t!]{.17\textwidth}
        \centering
        \includegraphics[scale=.35]{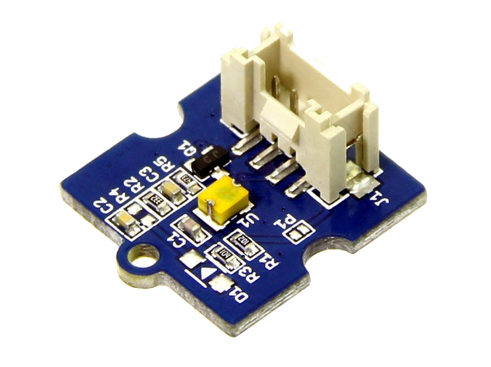}
        \caption{}
        \label{collisons}
    \end{subfigure}
    ~ 
    \begin{subfigure}[t!]{.17\textwidth}
        \centering
        \includegraphics[scale=.06]{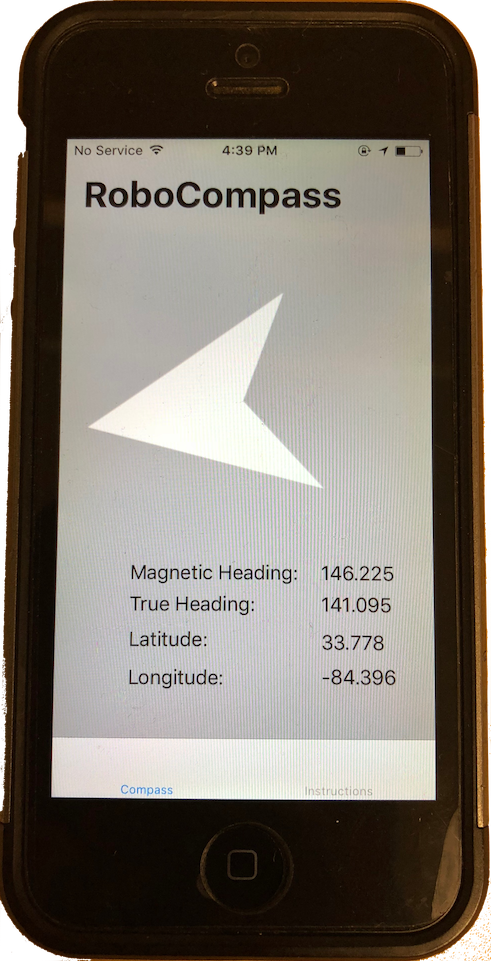}
        \caption{}
        \label{compass}
    \end{subfigure}
        ~ 
    \begin{subfigure}[t!]{.17\textwidth}
        \centering
        \includegraphics[scale=.35]{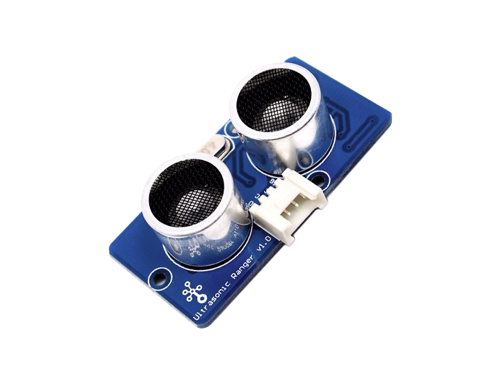}
        \caption{}
        \label{sonar}
    \end{subfigure}
            ~ 
    \begin{subfigure}[t!]{.17\textwidth}
        \centering
        \includegraphics[scale=.4]{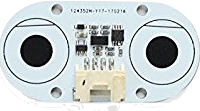}
        \caption{}
        \label{laserdist}
    \end{subfigure}
            ~ 
    \begin{subfigure}[t!]{.17\textwidth}
        \centering
        \includegraphics[scale=.15]{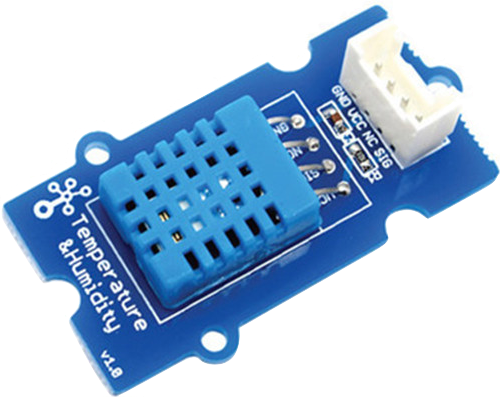}
        \caption{}
        \label{temp}
    \end{subfigure}
    \caption{Images of the HoneyBot sensors (a) Collision sensor (b) iPhone 5 Compass (c) Ultrasonic sonar (d) Laser Distance sensor (e) and a Temperature sensor.}
    \label{hbsensors}
\end{figure*}

The HoneyBot had five sensors, shown in Figure \ref{hbsensors}: a Sensolute MVS0608.02 Collision Sensor, an iPhone 5 Compass, a GrovePi SEN10737P Ultrasonic Sonar, a Dexter Industries Laser Distance Sensor, and an Aosong DHT11 Temperature Sensor. These sensors were chosen because of their significance to real-world ground robot applications. A collision sensor can be crucial to the well being of navigational robots, that are autonomous or remotely controlled, as they are the first line of defense for detecting and preventing costly damage to robotic end-effectors due to robot crashes \cite{ati-collision}. A collision sensor on a deployed autonomous robotic system can be used to report damage to relevant parties who may be physically distant. In order to know where to dispatch rescue teams, monitoring parties need to know the robots' location. This is where the compass, laser distance, and ultrasonic sonar sensors come in to play. While they don't provide absolute location, like a GPS would, in many indoor or well-defined environments they are equally useful.

An iPhone 5 was used as the compass for the HoneyBot because it provided more accurate headings than the STMicroelectronics LSM303D 6-Axis Accelerometer \& Compass. The small magnetometer in the accelerometer could not overcome the interference from the many electrical components on the GoPiGo 3 and produced inaccurate data. The iPhone 5 has much better internal component shielding and did not suffer from interference when placed near the robot. An iOS mobile application, called RoboCompass, was written in the Swift programming language and downloaded to the phone. The RoboCompass App sends compass readings in IP dataframes to the HoneyBot web server every time the compass reading changes (every time the robot moves).  

\section{HoneyBot Experimental Design}\label{experiment design}
\begin{figure}[b]
\centering
\includegraphics[scale=.2]{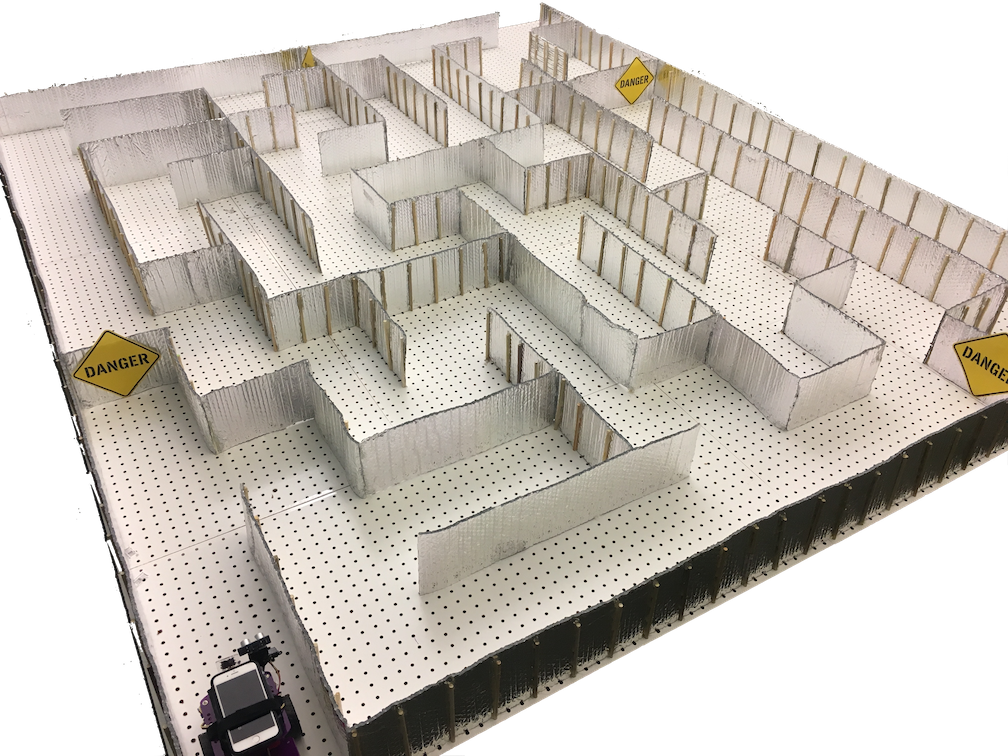}
\caption{Physical HoneyMaze with danger signs throughout.}
\label{honeymaze}
\end{figure}

Since the HoneyBot was built on a ground robot, the best form of evaluation was determined to be a navigational task. To support this, an evaluation arena was built in the form of a 10 x 12 foot maze (shown in Figure \ref{honeymaze}) and participants (with no prior knowledge of the research) were recruited over the course of one week to remotely navigate the HoneyBot through it. Before beginning this study IRB approval was requested from the Georgia Tech Office of Research Integrity Assurance and the experiment protocol was designed.

The "HoneyMaze" was constructed from approximately six 2 x 4 foot pegboards (used for the base or ground surface) and several hundred 1/2 x 48 inch wooden round dowels. The wooden dowels were cut, using circular saw equipment from the Georgia Tech ECE Senior Design Lab, into 6 inch pegs. These 6 inch pegs were then strategically "nailed" into the pegboard base, one peg per every 3 peg holes, in the design of the pre-selected maze. After the pegs were secured in place rolls of 48 inch x 25 ft reflective insulation were cut into 7 inch tall strips and hot glued to one side of the pegs to create barrier walls. Reflective insulation was used as the wall material because a positive correlation was identified between the robot's distance sensor accuracy and the reflectivity of the surfaces measured against. 

\begin{figure}[h]
\centering
\includegraphics[scale=.27]{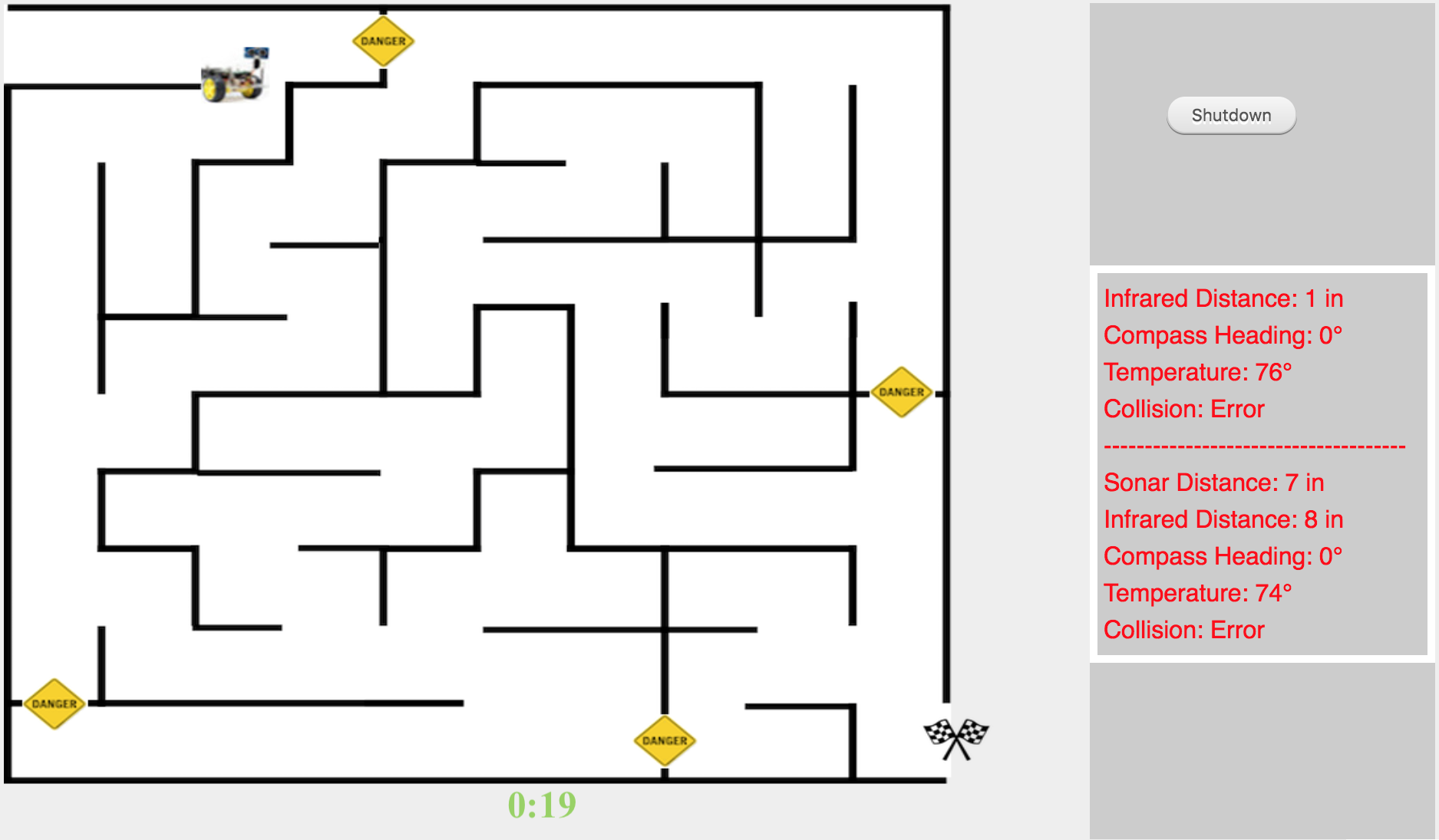}
\caption{HoneyBot user experiment website.}
\label{webpage}
\end{figure}

The experiment required individuals to read instructions on how to navigate a robot through an online maze and then access the robot through a web interface. The participants were told the online robot corresponded to a real robot who at their every arrow keystroke would actuate through a life sized maze identical to the one on their screen. Subjects were told that their mission was to navigate the robot through the maze fast as possible using only the online GUI maze and live sensor values from the robot displayed on screen. They were informed that the research objective was to determine the optimal constraint profile for the best performance and efficiency of the remotely controlled robot. A screenshot of the website is shown in Figure \ref{webpage}. At the bottom of the figure is a timer, participants were given 75 seconds to preview the maze and plan a route, then 60 seconds to actually navigate the robot to the finish flags. The "constraints" they were told was that the robot moves very slow and they should plan their routes wisely, making sure to consider all possible options. The research subjects were otherwise given no specific guidance concerning the danger signs and when asked about them experiment proctors only responded with "the meaning of the danger sign is up for interpretation, consider all possible options". To assist with the participant recruitment process and as an added incentive for subjects to strive to complete the maze quickly, they were promised \$5 for participating and \$10 if they completed the maze before their 60 seconds ran out. 

The real merit of the study, which is what the research participants were not told, is that the danger signs mark "shortcuts" through the online maze and the navigation task cannot be completed in the 60 second time limit without cutting through at least two of them. The danger signs can be thought of as the honey or vulnerable resource on a real system/network tempting attackers to compromise it. The real maze contains walls where the danger signs are located and if they decided to take the "shortcut", the real robot would stop actuating through the real maze and simulate all further interactions. After cutting through a danger sign the online GUI robot continues through the maze "normally", but all data output to the sensor control panel from that point on is spoofed. After completing the experiment participants are given a survey, and asked about the choices they made and what happened during the experiment.

\subsection{HoneyBot Sensor Model Development}\label{sensormodels}
Empirical observations were used to build the \textit{Device Models} for the HoneyBot sensors. For the temperature sensor, compass, laser distance sensor, and ultrasonic sonar the model development process was as follows:

\begin{enumerate}
\item The HoneyBot was placed at a viable maze location.
\item A Python script for sensor data collection was executed and given the robots' coordinates in the physical maze. After that the program polls the robots' sensors for data values.
\item The script then creates an index in a CSV file with the given coordinates and adds the sensor values to the index.
\end{enumerate}

This process is repeated several times at each of the 60 viable maze locations. A "viable maze location" is defined as an allowable maze location for the robot to navigate to. Once these models were built the collision sensor device model was very simple. Since the robot was not allowed to perform commands that could actually cause it to crash, the only time the collision sensor needed to read "True" was when the robot "cut through" a danger sign. To do this the actual reading from the collision sensor ("False") was always outputted to the user, unless they "cut through" a danger sign. At that point the collision sensor outputted "True" and the ultrasonic sonar/laser distance sensor outputted 0 for consistency. This was to really create the illusion that the robot hit an obstacle, but managed to keep going.

\section{Experiment Results}
The purpose of this experiment was to evaluate the HoneyBot and determine how convincing the Sensor \textit{Device Models} developed from real observations were. Of particular interest in the study were participants who "cut through" danger signs to complete the maze quicker, because that action automatically triggered the \textit{Input Verification Module} of the Robot Controller, which stopped the  real robot from actuating and initiated the simulation.

\subsection{Research Subject Demographics and Statistics}
\begin{figure}[h]
\centering
\includegraphics[scale=.3]{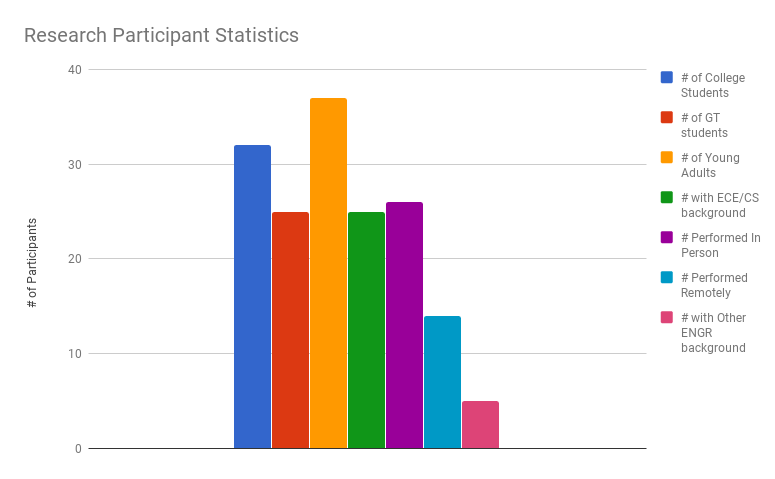}
\caption{HoneyBot user experiment research participant statistics.}
\label{subjectstats}
\end{figure}

\begin{figure}[h]
\centering
\includegraphics[scale=.3]{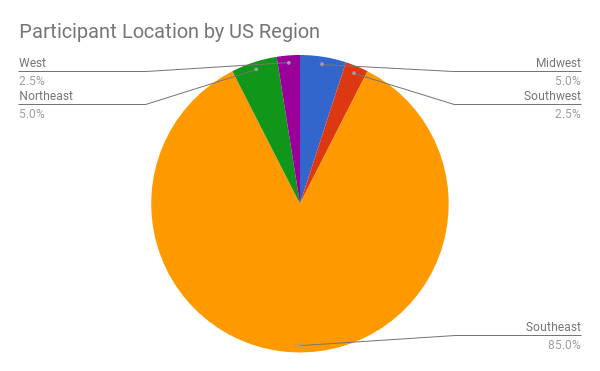}
\caption{HoneyBot user experiment research participant locations by US region.}
\label{regionstats}
\end{figure}

\begin{figure}[h]
\centering
\includegraphics[scale=.3]{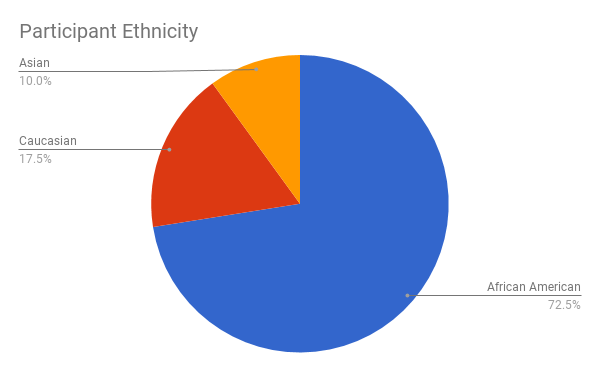}
\caption{HoneyBot user experiment research participant ethnicity breakdown.}
\label{racestats}
\end{figure}

The research experiments took place on the Georgia Tech Atlanta Campus over the course of one week and was performed by 40 individuals from various academic/cultural backgrounds, physical locations across the US, and stages of life. The vast majority of subjects (95\%) were young adults between the ages of 18 and 26. Figures \ref{subjectstats}, \ref{regionstats}, and \ref{racestats} give some quick statistics about the research subjects, including their regional location and cultural background.

\subsection{HoneyBot Experiment Findings}
The HoneyBot experiment took approximately 20 minutes per participant and consisted of an initial instruction overview, the actual experiment completion, and a concluding survey. The Robot Experiment Survey was distributed through Qualtrics Online Survey Software, and consisted of, at most 12 questions. Certain questions were displayed/omitted based on participant responses. For example, if Question 4, 'was the overall experiment completion process difficult?', was answered 'No' then Question 5, 'what made it difficult?', was not asked. Five questions on the other hand, were always asked. They served to provide baseline knowledge about the subjects' experience. The questions were:

\begin{enumerate}
\item Were you able to navigate from start to finish of the maze within the time limit? (Y/N)
\item Map your navigated route by selecting the letters on the graphic below. If you did not finish the maze select to the nearest point you reached. (A-Z)
\item On a scale of 1-5, where 1 is very inaccurate and 5 is very accurate, how accurate did the sensor values displayed on the control panel seem throughout the experiment? (1-5)
\item Was the overall experiment completion process difficult? (Y/N)
\item Did you at any point cross through a danger sign? (Y/N)
\end{enumerate} 

\begin{figure*}[h]
\centering
\includegraphics[scale=.4]{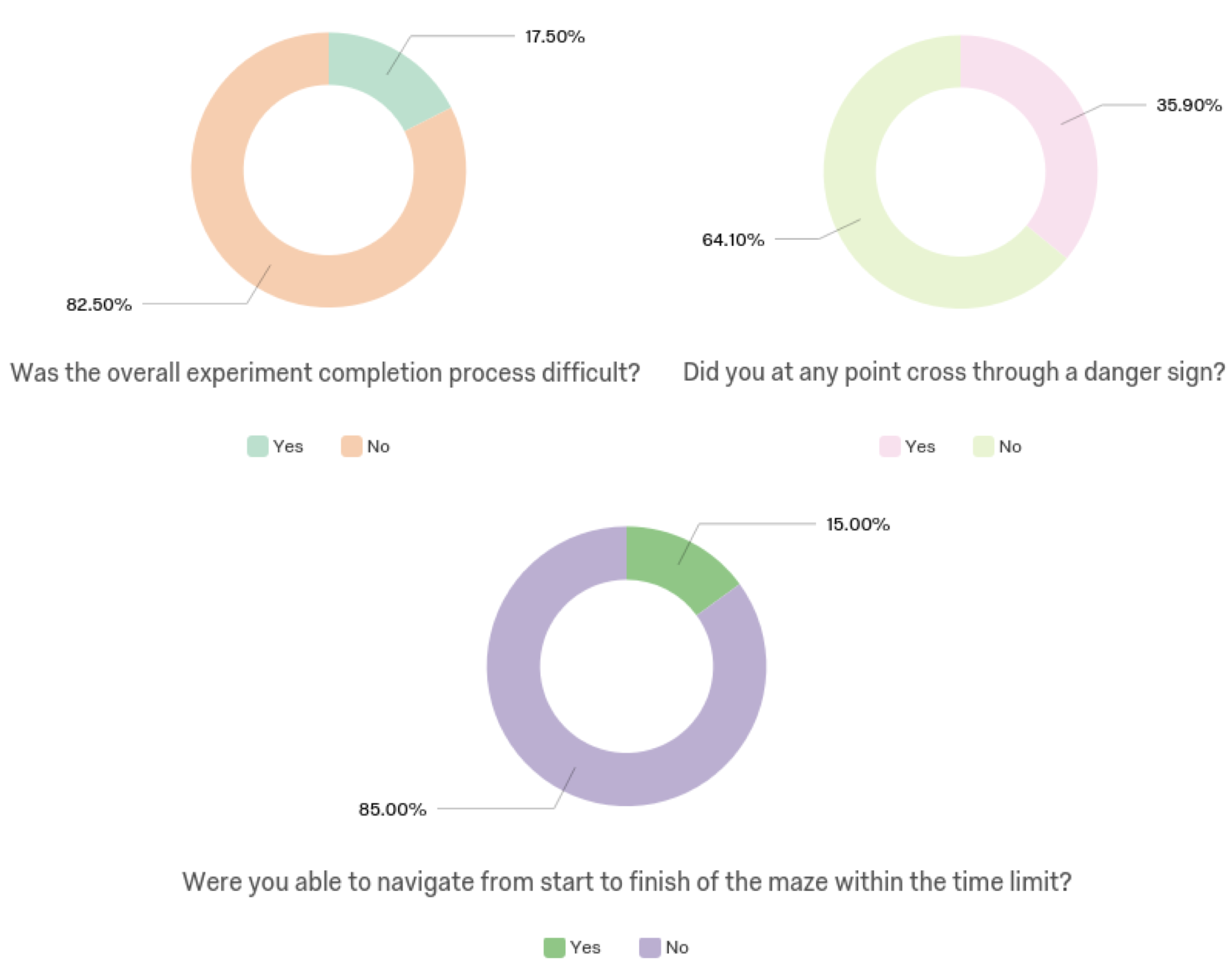}
\caption{Survey responses to Question 4, Question 6, and Question 1.}
\label{q1-q4-q6}
\end{figure*}

Figure \ref{q1-q4-q6} shows the user responses to survey questions 1, 4, and 6. It can be gathered from the pie charts that the overall experiment completion process was not difficult, most people did not finish the maze in the allotted time, and a little over a third of the participants (14 people) "cut through" at least one danger sign and were shown simulated sensor values. 

\begin{figure}[ht!]
    \centering
    \begin{subfigure}{.10\textwidth}
        \centering
        \hspace*{-.5cm}
        \includegraphics[scale=.11]{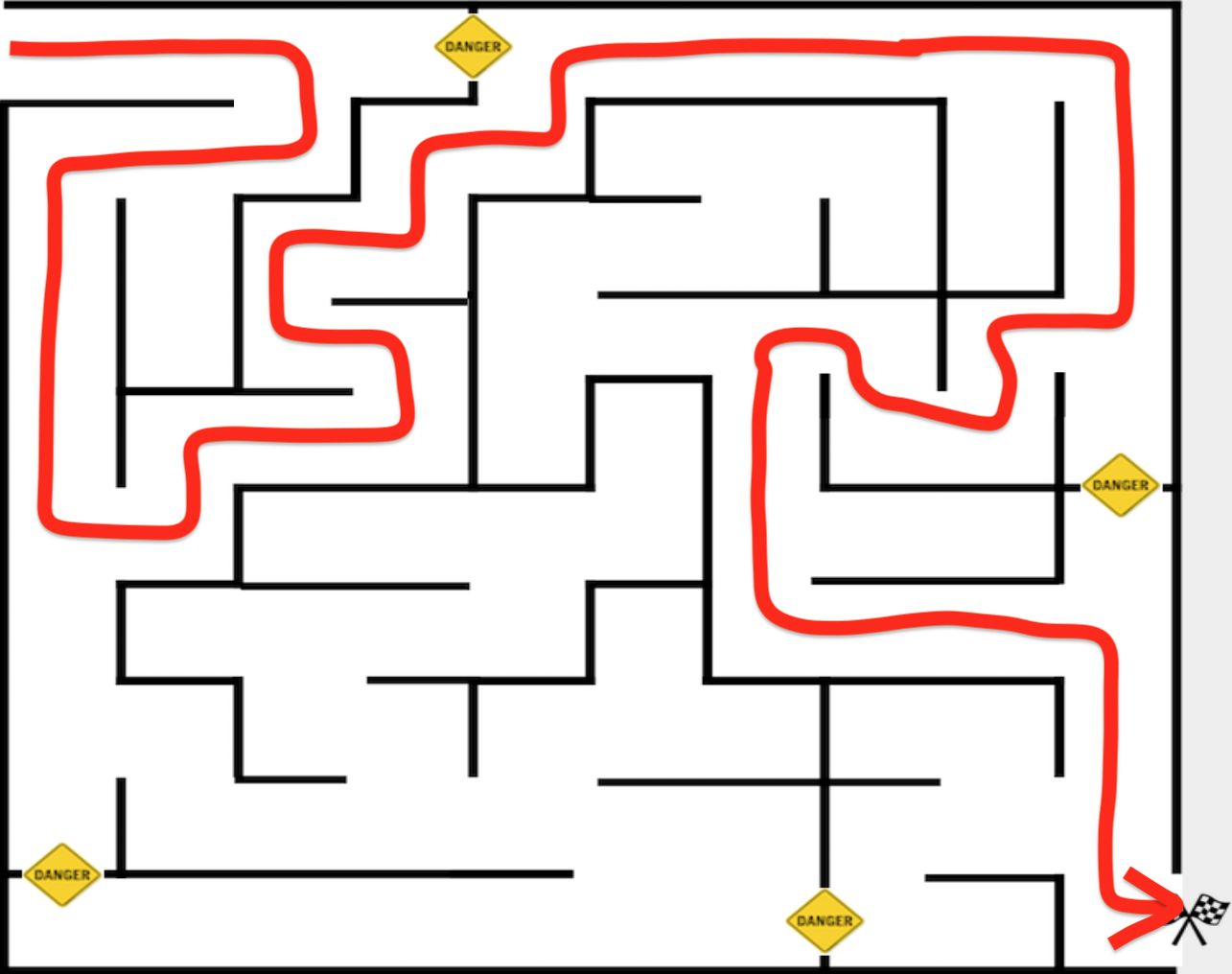}
        \caption{}
        \label{route1}
    \end{subfigure}
    ~ 
    \begin{subfigure}{.10\textwidth}
        \centering
        \includegraphics[scale=.11]{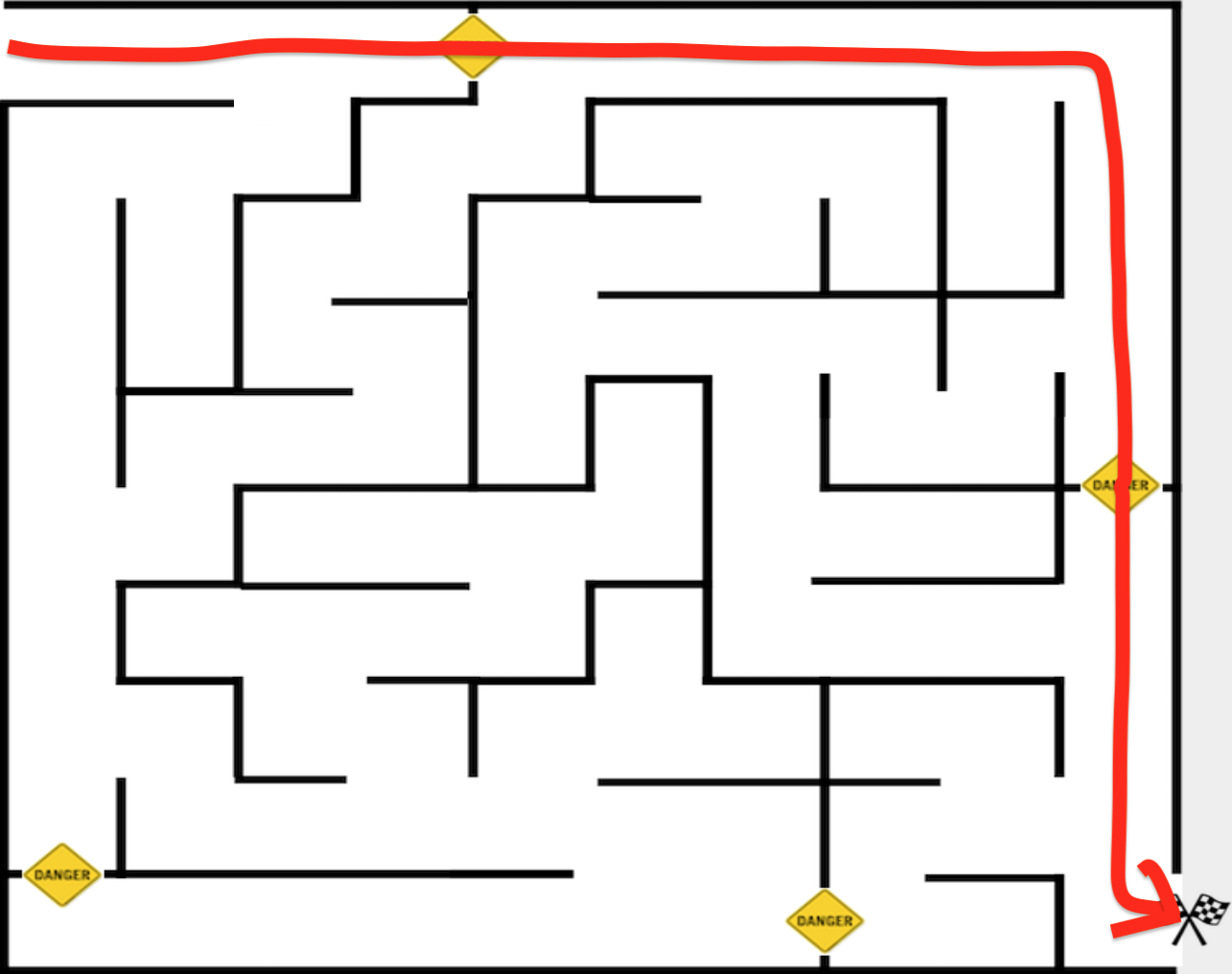}
        \caption{}
        \label{route2}
    \end{subfigure}
        ~ 
    \begin{subfigure}{.10\textwidth}
        \centering
        \hspace*{.35cm}
        \includegraphics[scale=.11]{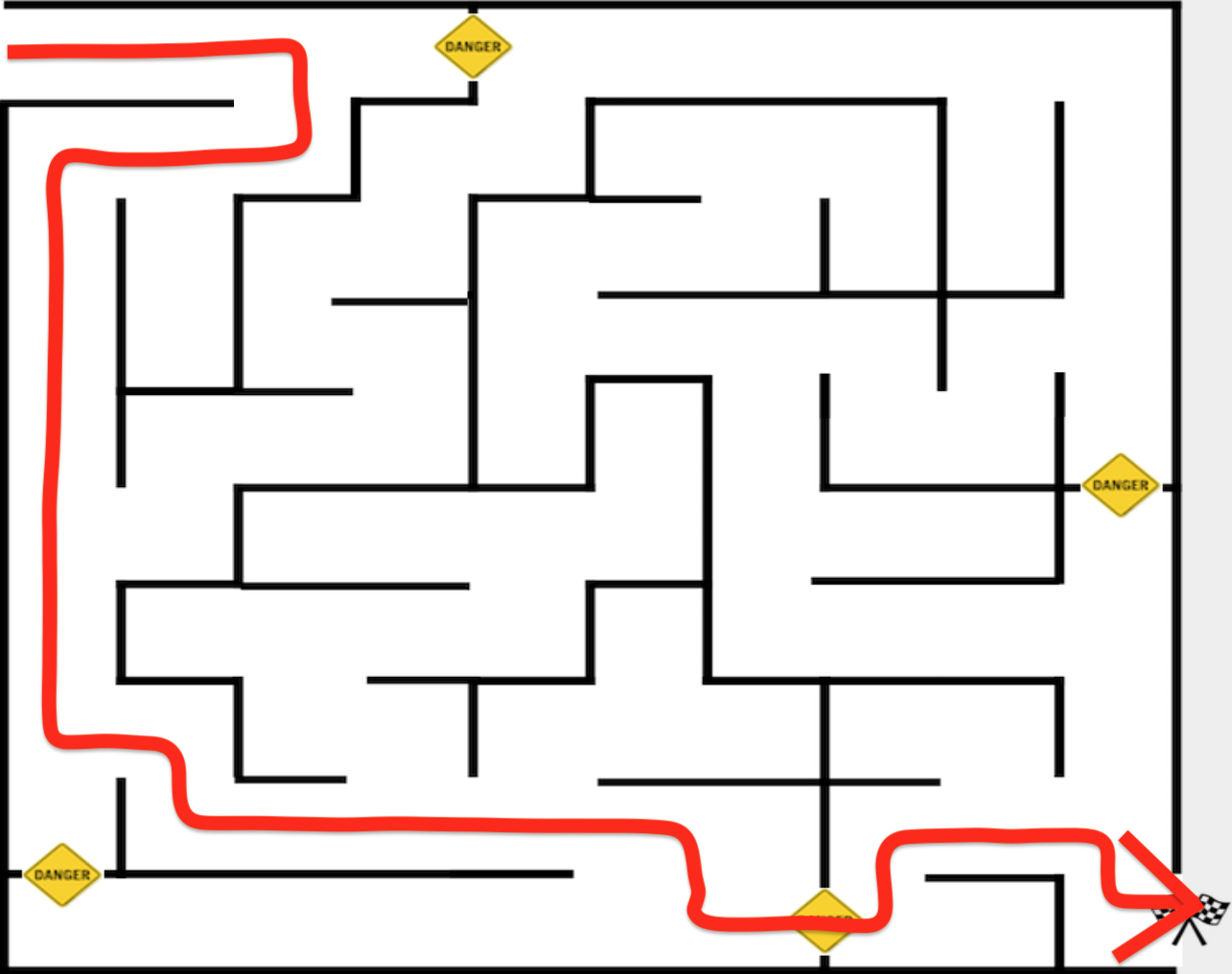}
        \caption{}
        \label{route3}
    \end{subfigure}
    \caption{Top three most navigated routes by participants (a) 56\% of subjects took this route (b) 22\% of subjects took this route (c) and 1\% of participants took this route.}
    \label{top3routes}
\end{figure}

Question 2 came with the labeled maze, and according to the survey results, two of the three most traveled paths "cut through" danger signs. 79\% of subjects took the top three most navigated paths and the only other consistently navigated path (taken by 3 participants) also "cut through" a danger sign. It is important to note that the routes depicted in Figure \ref{top3routes} indicate "attempted" navigated routes, most subjects did not complete the maze, but indicated that was the route they intended to take.

In total, 14 participants cut through danger signs and triggered the HoneyBot 'simulation mode'. Table \ref{table2} shows that of all 40 participants surveyed, 70\% of them rated the sensor accuracy during the whole experiment a 3 or 4 (mean of 3.58) out of 5. And of the 14 participants who cut through a danger sign and unknowingly experienced simulated sensor values, 71\% rated the sensor accuracy a 4 or 5 (mean of 3.86) out of 5. This can be interpreted to mean research subjects did not notice a difference between the simulated sensor values and the real sensor values coming from the HoneyBot. From this it can be concluded that the HoneyBot developed successfully fools "deviant users" and the Sensor \textit{Device Models} effectively mirror reality.


\section{Conclusion}
The need for security in the field of robotics is growing and will continue growing as robots become ubiquitously integrated into everyday life. Networked systems will always be vulnerable and susceptible to exploits, but safeguards should be put in place to ensure:

\begin{itemize}
\item Robotic systems able to distinguish between safe and unsafe actions they are commanded to perform
\item The system uses this distinction to protect itself from physical harm
\item There are reliable mechanisms means for system administrators to learn of compromise
\item There are methods for monitoring system intruders
\end{itemize}

As a proposed solution to these problems our previous work introduced the HoneyBot \cite{celine}, the first honeypot specifically designed for robotic systems. The HoneyBot uses techniques from traditional honeypots and \textit{Device Models} built for common robotic sensors to simulate unsafe actions and physically perform safe actions to fool attackers. Unassuming attackers are led to believe they are connected to an ordinary robotic system, believing their exploits are being successfully executed. All the while the HoneyBot is logging all communications and exploits sent to be used for attacker attribution and threat model creation. In this paper, we presented the results of a user experiment performed to show the feasibility of the HoneyBot framework and architecture as it applies to real robotic systems and found that on average research subjects could not differentiate between simulated sensor values and the real sensor values coming from the HoneyBot. From this, we conclude that the HoneyBot developed successfully fools "deviant users" and the Sensor \textit{Device Models} effectively mirror reality.

\section{Future Work}\label{future}
The overarching goal of this research was to evaluate the effectiveness a robotic system that could reasonably convince remote connected attackers with malintent that their malicious payloads are successful, while in reality simulating data responses and preserving the real system. This was done through a user study with the HoneyBot built on the GoPiGo 3 platform. While the preliminary results of the research study are promising, there is more work that can be done to improve  the robustness of the ideas and implementations presented.

\begin{table*}[h]
\centering
\caption{Survey results to questions about robot sensor accuracy}
\begin{adjustbox}{width=\textwidth} 
\label{table2}
\begin{tabular}{@{}ccc@{}}
\toprule
\begin{tabular}[c]{@{}c@{}}Scale \\ (1 is very inaccurate and 5 is very accurate)\end{tabular} & \multicolumn{1}{l}{\begin{tabular}[c]{@{}l@{}}How accurate did the sensor values \\ displayed on the control panel seem \\ throughout the experiment?\end{tabular}} & \multicolumn{1}{l}{\begin{tabular}[c]{@{}l@{}}How accurate did the sensor values \\ displayed on the control panel seem \\ after you crossed through danger sign(s)?\end{tabular}} \\ \midrule
1 & 1 (2.5\%) & 1 (7.14\%) \\
2 & 4 (10\%) & 0 (0\%) \\
3 & 13 (32.5\%) & 3 (21.43\%) \\
4 & 15 (37.5\%) & 6 (42.86\%) \\
5 & 7 (17.5\%) & 4 (28.57\%) \\ \midrule
Total & 40 (100\%) & 14 (100\%) \\ \bottomrule
\end{tabular}%
\end{adjustbox}
\end{table*}

\subsection{Evaluation Caveats}
 The robot experiment evaluation, though it proved the HoneyBot was convincing should be taken with a grain of salt. The small sample size of 40, is not enough to draw far-reaching conclusions. More user testing needs to be done to solidify the preliminary conclusions drawn. In addition to this, while there were some safeguarding (research proctors monitored the experiment task) against user falsifying the self reported survey results, there is always the possibility of fabrication when human subjects are involved and this must be considered. 
 
\subsection{Rethinking HoneyBot Remote Access Mechanisms and Evaluation Redesign}
One possible future direction for this work is a change in HoneyBot remote access techniques and new methods for evaluation. For the user study the HoneyBot was accessible through a website which was functional, but had few usability issues. Another more reliable mechanism for system accesses which would aid in evaluation could be via a command line tool such as SSH or even a graphical VNC. Otherwise, if a website is the medium of choice, to reduce lag and support a multi-user evaluation it is necessary to run the web server securely off-site. And the server should have enough resources to handle many web requests simultaneously.  

Future evaluations of the HoneyBot should not only involve human user testing, but general performance metrics as well. It would be important to note differences in response times of real vs simulated responses sent over the network, as well as any detectable footprint the HoneyBot software would leave on the system. In order to remain undetectable, processes should be hidden or run in an obfuscated manner very similar to a root-kit. Attackers/malicious parties who somehow gain full access to the system should not in theory be able to notice a difference between identical systems if one is running the HoneyBot software and the other is not.

In an effort to welcome contributions and continuations of this research we have published source code for the HoneyBot on GitHub as well as documents used for the user study. The HoneyBot iPhone Compass App source code can be found on GitHub at  \href{https://github.gatech.edu/cirvene3/HoneyBot/tree/master/Swift/RoboCompass}{RoboCompass Code}. The instructions given to study participants before the experiment can be found on GitHub at  \href{https://github.gatech.edu/cirvene3/HoneyBot/blob/master/RobotExperimentInstructions.pdf}{User Experiment Instructions}. Finally, the full survey participants completed after the experiment can be found on GitHub at  \href{https://github.gatech.edu/cirvene3/HoneyBot/blob/master/Qualtrics_Survey.pdf}{Qualtrics Survey}.








\section*{Acknowledgment}
  The authors would also like to thank the anonymous referees for
  their valuable comments and helpful suggestions. The work is
  supported by the National Science Foundation under Grant
  No.:{1544332}.

\bibliographystyle{IEEEtranS}
\bibliography{bibliography}

\end{document}